\documentclass[12pt]{iopart}
\usepackage{graphicx}
\usepackage{epsfig}
\usepackage{dcolumn}
\usepackage{bm}
\usepackage[usenames]{color}
\usepackage{amssymb}
\usepackage{setstack,iopams}
\begin{document}
\title[Analytical approach on linear and nonlinear pulse]{Analytical approach on linear and nonlinear pulse propagations
in an open $\Lambda$-type molecular system with Doppler broadening}

\author{Chaohua Tan, Chengjie Zhu and Guoxiang Huang$^1$}
\address{State Key Laboratory of Precision Spectroscopy and
Department of Physics, East China Normal University, Shanghai
200062, China}
\ead{gxhuang@phy.ecnu.edu.cn}

\date{\today}

\begin{abstract}
We develop a systematic analytical approach on linear and nonlinear
pulse propagations in an open $\Lambda$-type molecular system with
Doppler broadening. In linear case, by using residue theorem and a
spectrum decomposition method, we prove that there exists a
crossover from electromagnetically induced transparency (EIT) to
Autler-Townes splitting (ATS) for co-propagating configuration of
probe and control fields. However, there is no EIT and hence no
EIT-ATS crossover for counter-propagating configuration. We give
various explicit formulas, including probe-field spectrum
decomposition, EIT condition, width of EIT transparency window, as
well as a comparison with the result of cold molecules. Our
analytical result agrees well with the experimental one reported
recently by A. Lazoudis {\it et al}. [{\it Phys. Rev.} A {\bf 82},
023812 (2010)]. In nonlinear case, by using the method of
multiple-scales, we derive a nonlinear envelope equation for
probe-field propagation. We show that stable ultraslow solitons can
be realized in the open molecular system.

\end{abstract}


\submitto{\jpb}

\maketitle

\section{INTRODUCTION}{\label{Sec:1}}

In recent years, much attention has been paid to the study of quantum coherent phenomena
in various multi-level systems, typical examples include Auter-Townes splitting (ATS)~\cite{at}
and electromagnetically induced transparency (EIT)~\cite{fle}. Such phenomena are
not only important from viewpoint of basic research, but also very attractive for many practical
applications, such as lasing without inversion, coherent population transfer, enhanced Kerr
nonlinearity, slow light, quantum memory,  atom and/or photon entanglement, precision spectroscopy,
precision measurement, and so on~\cite{fle,khu}.

ATS occurs when absorption spectrum of a quantum transition can be decomposed into a sum of two net Lorentzian terms if one of two levels involved in the transition is coupled to a third level induced by a strong control field. EIT occurs when the absorption spectrum can be decomposed not only into two Lorentzians, but also with additional quantum {\it destructive} interference term(s). Usually, in systems with ATS or EIT, a transparency window is opened. However, the opening of the transparency window cannot be tell us whether the phenomenon belongs to ATS or EIT, each of which has different physical origin. ATS happens only for strong control field, but EIT happens even the control field is weak. Especially, Only for weak control field can essential characters of EIT be illustrated
clearly~\cite{Agarwal1997,Anisimov2008,Tony2010,Anisimov2011}.

EIT in various atomic systems has been studied intensively both theoretically and
experimentally~\cite{fle,khu}. However, systematic investigations of EIT in molecular systems are still lacking. Up to now, there are only several related experimental studies in molecular systems, including the
EIT in $^7$Li$_2$~\cite{Qi2002}, K$_2$ \cite{Li2005} and Na$_2$ vapors~\cite{Lazoudis2008,Lazoudis2011}, in  acetylene molecules filled in hollow-core photonic crystal fibers~\cite{Ghosh2005,Benabid2011} and in photonic microcells~\cite{Light2009}, and in Cs$_2$ in a vapor cell~\cite{Li10}, and so on. Major difficulties for observing EIT in molecules are small transition-dipole-moment matrix elements in comparison with those in atoms, and many decay pathways to other molecular states not involved in the main excitation scheme.

In an interesting work reported recently by Lazoudis {\it et al.}~\cite{Lazoudis2010}, EIT in an open hot $\Lambda$-type molecular $^7$Li$_2$ system has been studied experimentally.  A numerical simulation under steady-state approximation is used by the authors for solving density matrix equations for molecules. Though the numerical simulation is helpful to explain experimental data, it is however hard to discern ATS from EIT objectively because the physical mechanism behind numerical results are not clear. In particular, since open molecular systems with Doppler broadening are very complicated and have very different features in different parameter regions, it is necessary to clarify in an analytical way the quantum interference characters inherent in such systems, which, to the best of our knowledge, has not been done in literature up to now. In addition, it is also necessary to go beyond steady-state approximation if probe pulse is used in experiment.

In this work, we develop a systematic analytical approach on linear and nonlinear pulse propagations in open $\Lambda$-type molecular systems with Doppler broadening.  In linear case, by using residue theorem and spectrum decomposition method,  we prove clearly that there exists a crossover EIT to ATS for co-propagating configuration of probe and control fields. However, there is no EIT and hence no EIT-ATS crossover for counter-propagating configuration. We provide various explicit formulas, including probe-spectrum decomposition, EIT condition, and width of EIT transparency window, as well as a comparison
with the result of cold molecules. Our analytical result agrees well with the experimental one reported recently by A. Lazoudis {\it et al}. \cite{Lazoudis2010}. In nonlinear case, by using a standard method of multiple-scales, we derive a nonlinear envelope equation for probe-field propagation. We show that a stable ultraslow solitons can be realized in the open molecular system. Notice that  nonlinear pulse propagation in coherent atomic systems via EIT has attracted tremendous attention in recent
years~\cite{Hong2003,Wu2004,Huang2005,Hang2006,Michinel2006,Huang2008,Yang2010,LiHuang2010}, nobody however has considered similar problem for molecules till now.

The article is arranged as follows. In the next section we present
our model and associated Maxwell-Bloch (MB) equations. In
section~\ref{Sec:hot_m}, we consider the linear property of the
system by using residue theorem and spectrum decomposition method.
Quantum interference characters for hot molecules with both co- and
counter-propagating configurations and also for cold molecules are
analyzed in detail.  In section~\ref{sec_soliton}, the method of
multiple-scales is used to study the weak nonlinear propagation of
the probe field.  Lastly, section~\ref{sec_conclusion} contains a
summary of the main results obtained in our work.

\section{Model}{\label{Sec:2}}

\begin{figure}
\centering
\includegraphics[scale=0.4]{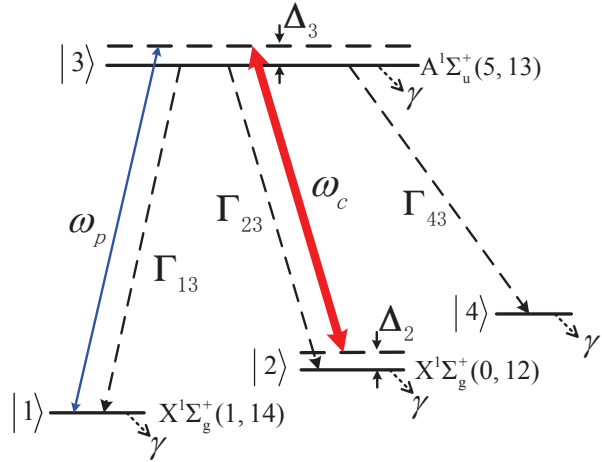}\\
\caption{(Color online) $\Lambda$-type EIT scheme for open Li$_2$
molecular system. Excited state A$^1\Sigma_u^+(v'=5,J'=13)$ (labeled
$|3\rangle$) couples to  ground state X$^1\Sigma_g^+(v''=0,J''=12)$
(labeled $|2\rangle$) by the control field with center frequency
$\omega_c$ and also to another ground state
X$^1\Sigma_g^+(v''=1,J''=14)$ (labeled $|1\rangle$) by the probe
field with center frequency $\omega_p$. $\Delta_{2}$ and
$\Delta_{3}$ are detunings, $\Gamma_{jl}$ are population decay rates
from $|l\rangle$ to $|j\rangle$, and $\gamma$  is transit rate.
Molecules occupying the excited state $|3\rangle$ may decay to many
other states besides the states $|1\rangle$ and $|2\rangle$. All
these other states are represented by state
$|4\rangle$.}\label{mod1}
\end{figure}

The model adopted here is the same as that used in~\cite{Lazoudis2010}.
An open three-state $\Lambda$-type Li$_2$
molecular system (figure~\ref{mod1})
%
%
consists of an exited upper-level A$^1\Sigma_u^+(v'=5,J'=13)$
(labeled $|3\rangle$) and two ground states
X$^1\Sigma_g^+(v''=1,J''=14)$ (labeled $|1\rangle$) and
X$^1\Sigma_g^+(v''=0,J''=12)$ (labeled $|2\rangle$). A control
field with center frequency $\omega_c$ couples to the excited state
$|3\rangle$ and the ground state $|2\rangle$. The other ground state
$|1\rangle$ couples to the $|3\rangle$ by a
probe field  with center frequency $\omega_p$.
The exited level $|3\rangle$ decay spontaneously
to the ground states $|1\rangle$ and $|2\rangle$ with decay rates $\Gamma_{13}$
and $\Gamma_{23}$, respectively. The parameter $\gamma$ represents
the transient relaxation rate of the molecule entering and leaving
interaction region between light and the molecule. It reflects also
the additional relaxation of each state due to the interaction with
thermal reservoir~\cite{Lazoudis2010}. The electric field vector of the system is
$\mathbf{E}=\sum_{l=p,c}\mathbf{e}_l{\cal
E}_l(z,t)e^{i(\mathbf{k}_l\cdot\mathbf{r}-\omega_lt)}+$c.c., where
$\mathbf{e}_l$ $(\mathbf{k}_l)$ is the unit polarization vector (wave
number) of the electric field component with the envelope ${\cal
E}_l$ $(l=p,c)$.

As indicated in the last section, decay processes in molecular
systems are very complicated in comparison with those of atoms.
There exist many decay pathways to other molecular states not
involved in the main excitation scheme, and hence the theoretical
model considered is necessarily an open one. In the excitation
scheme adopted above, molecules occupying the excited level
$|3\rangle$ may follow various relaxation pathways and decay to many
lower vibration-rotation levels besides the levels $|1\rangle$ and
$|2\rangle$. In our modeling all these levels are represented by the
level $|4\rangle$. The decay rate $\Gamma_{43}$ indicates the
spontaneous emission rate of level $|3\rangle$ to level $|4\rangle$
(see figure~\ref{mod1}).

For hot molecules, inhomogeneous Doppler broadening must
be taken into account because the experiments are carried out in a heat-pipe
oven~\cite{Lazoudis2010}. The Hamiltonian of the system in interaction picture
under electric-dipole and rotating-wave approximations is
\begin{eqnarray}
\hat{H}=-\hbar(\Omega_ce^{i[\mathbf{k}_c\cdot(\mathbf{r}+\mathbf{v}t)
-\omega_ct]}|3\rangle\langle2|+\Omega_pe^{i[\mathbf{k}_p\cdot(\mathbf{r}
+\mathbf{v}t)-\omega_pt]}|3\rangle\langle1|+{\rm c.c.}),\label{H}
\end{eqnarray}
where $\bf{v}$ is molecular velocity, $\Omega_{c(p)}=\left( {\bf
e}_{c(p)}\cdot\boldsymbol{\mu}_{32(31)}\right) {\cal
E}_{c(p)}/(2\hbar)$ is half Rabi frequency of the control (probe)
field, with $\boldsymbol{\mu}_{jl}$ the electric-dipole matrix
element associated with the transition from state $|j\rangle$ to
state $|l\rangle$.
%
The optical Bloch equation in the interaction picture reads
\begin{eqnarray}
&&i\frac{\partial}{\partial
t}\sigma_{11}+i\gamma(\sigma_{11}-\sigma_{11}^{{\rm eq}})-i\Gamma_{13}\sigma_{33}+\Omega_{p}^{*}\sigma_{31}-\Omega_{p}\sigma_{31}^{*}=0,\nonumber\\
&&i\frac{\partial}{\partial
t}\sigma_{22}+i\gamma(\sigma_{22}-\sigma_{22}^{{\rm
eq}})-i\Gamma_{23}\sigma_{33}+\Omega_{c}^{*}\sigma_{32}-\Omega_{c}\sigma_{32}^{*}=0,\nonumber\\
&&i\frac{\partial}{\partial
t}\sigma_{33}+i\gamma(\sigma_{33}-\sigma_{33}^{{\rm
eq}})+i\Gamma_{3}\sigma_{33}
+\Omega_{p}\sigma_{31}^{*}+\Omega_{c}\sigma_{32}^{*}\nonumber\\
& & -\Omega_{p}^{*}\sigma_{31}-\Omega_{c}^{*}\sigma_{32}=0,\nonumber\\
&&i\frac{\partial}{\partial
t}\sigma_{44}+i\gamma(\sigma_{44}-\sigma_{44}^{{\rm
eq}})-i\Gamma_{43}\sigma_{33}=0,\nonumber\\
&&\left(i\frac{\partial}{\partial
t}+d_{21}\right)\sigma_{21}+\Omega_{c}^{*}\sigma_{31}-\Omega_{p}\sigma_{32}^{*}=0,\nonumber\\
&&\left(i\frac{\partial}{\partial
t}+d_{31}\right)\sigma_{31}+\Omega_{p}(\sigma_{11}-\sigma_{33})+\Omega_{c}\sigma_{21}=0, \nonumber\\
&&\left(i\frac{\partial}{\partial
t}+d_{32}\right)\sigma_{32}+\Omega_{c}(\sigma_{22}-\sigma_{33})+\Omega_{p}\sigma_{21}^{*}=0,\label{dme}
\end{eqnarray}
for nondiagonal elements, where
$d_{21}=-(\mathbf{k}_p-\mathbf{k}_c)\cdot\mathbf{v}+\Delta_{2}-\Delta_{1}+i\gamma_{21}$,
$d_{31}=-\mathbf{k}_p\cdot\mathbf{v}+\Delta_{3}-\Delta_{1}+i\gamma_{31}$,
$d_{32}=-\mathbf{k}_c\cdot\mathbf{v}+\Delta_{3}-\Delta_{2}+i\gamma_{32}$
with $\gamma_{jl}=(\Gamma_{j}+\Gamma_{l})/2+\gamma+\gamma_{jl}^{{\rm
col}}\ (j,l=1,2,3)$. Here $\Delta_j\ (j=1,2,3)$ are detunings,
and
$\Gamma_{j}$ denotes the total decay rate of population out of level
$|j\rangle$, which is defined by $\Gamma_{j}=\sum_{l\neq
j}\Gamma_{lj}$. The quantity $\gamma_{jl}^{{\rm col}}$ is the
dephasing rate due to processes such as elastic collisions.
$\sigma_{jj}^{\rm eq}$ is the thermal equilibrium value of
$\sigma_{jj}$ when all electric-fields are absent. Equation
(\ref{dme})  satisfies $\sum_{j=1}^4 \sigma_{jj}=1$ with
$\sum_{j=1}^4 \sigma_{jj}^{\rm eq}=1$. At thermal equilibrium,
population in the excited state $|3\rangle$ is much smaller than
that of the ground states, i.e. $\sigma_{33}^{\rm eq}\simeq 0$ and
hence $\sigma_{11}^{\rm eq}+\sigma_{22}^{\rm eq}+\sigma_{44}^{\rm
eq}\simeq 1$.

The evolution of the electric field is governed by the Maxwell equation.
%
Due to the Doppler effect, the electric polarization intensity of
the system is given by ${\bf P}={\cal N}_a\int_{-\infty}^{\infty}dv
f(v)\{ \boldsymbol{\mu}_{13}\sigma_{31} {\rm exp}[i(k_p z-\omega_p
t)]+\boldsymbol{\mu}_{23}\sigma_{32} {\rm exp}[i(k_c z-\omega_c
t)]+{\rm c.c.}\},$ where ${\cal N}_a$ is molecular density and
$f(v)$ is the molecular velocity distribution function. For
simplicity, we have assumed electric-field wavevectors are along
$z$-direction, i.e. ${\bf k}_{p,c}=(0,0,k_{p,c})$. Under the
slowly-varying envelope approximation, the Maxwell equation reduces
into
\begin{equation}\label{eqs:maxwell}
i\left(\frac{\partial}{\partial
z}+\frac{1}{c}\frac{\partial}{\partial t}\right)\Omega_p
+\kappa_{13}\int_{-\infty}^{\infty}dv f(v)\sigma_{31}(z,v,t)=0,
\end{equation}
with $\kappa_{13}={\cal
N}_a\omega_p|\boldsymbol{\mu}_{31}|^2/(2\hbar\varepsilon_0 c)$, here
$c$ is the light speed in vacuum.

The MB equations (\ref{dme}) and (\ref{eqs:maxwell}) are
our starting point for the study of linear and nonlinear pulse
propagations in the open molecular system with Doppler broadening.

\section{Linear propagation}\label{Sec:hot_m}

\subsection{Base state and general linear solution}

We first consider linear propagation of the probe field. For this
aim, one must know the base state $\sigma_{jl}^{(0)}$, i.e. the
steady-state solution of the MB equations (\ref{dme}) and
(\ref{eqs:maxwell}) for $\Omega_p=0$. It is easy to obtain
\begin{eqnarray}
&&\sigma_{11}^{(0)}=\frac{[\gamma\Gamma_{3\gamma}X_1+(2\gamma+\Gamma_{43})|\Omega_c|^2]\sigma_{11}^{{\rm
eq}}+\Gamma_{13}|\Omega_c|^2(1-\sigma_{44}^{{\rm
eq}})}{X_2},\nonumber\\
&&\sigma_{22}^{(0)}=\frac{\gamma[\Gamma_{3\gamma}X_1+|\Omega_c|^2]\sigma_{22}^{{\rm
eq}}}{X_2},\nonumber\\
&&\sigma_{33}^{(0)}=\frac{\gamma|\Omega_c|^2\sigma_{22}^{{\rm
eq}}}{X_2},\nonumber\\
&&\sigma_{44}^{(0)}=\frac{[\gamma\Gamma_{3\gamma}X_1+(2\gamma+\Gamma_{13})|\Omega_c|^2]\sigma_{44}^{{\rm
eq}}+\Gamma_{43}|\Omega_c|^2(1-\sigma_{11}^{{\rm
eq}})}{X_2},\nonumber\\
&&\sigma_{32}^{(0)}=-\frac{\Omega_c}{d_{32}}\cdot\frac{\gamma\Gamma_{3\gamma}X_1\sigma_{22}^{{\rm
eq}}}{X_2}\label{BS}
\end{eqnarray}
and $\sigma^{(0)}_{21}=\sigma^{(0)}_{31}=0$, where
$\Gamma_{3\gamma}\equiv\gamma+\Gamma_3$,
$X_1\equiv\{[(\Delta_3-\Delta_2)-k_cv]^2+\gamma_{32}^2\}/(2\gamma_{32})$
and $X_2\equiv
\gamma(\gamma+\Gamma_3)X_1+(2\gamma+\Gamma_{13}+\Gamma_{43})|\Omega_c|^2$.
Note that in above expressions $d_{21}=d_{21}(v)=-(k_p-k_c)
v+\Delta_{2}-\Delta_{1}+i\gamma_{21}$, $d_{31}=d_{31}(v)=-k_p
v+\Delta_{3}-\Delta_{1}+i\gamma_{31}$, and
$d_{32}=d_{32}(v)=-k_cv+\Delta_{3}-\Delta_{2}+i\gamma_{32}$.

When switching on the probe field, the base state (\ref{BS}) will be modified.
In linear theory, $\Omega_p$ is taken as a very small quantity.
At first order in $\Omega_p$, the populations and the coherence between the
states $|2\rangle$ and $|3\rangle$ are not changed, but with
\begin{eqnarray}
\Omega_p^{(1)}&&=F\,e^{i\theta},\nonumber\\
\sigma_{21}^{(1)}&&=-\frac{(\omega+d_{31})\sigma_{32}^{*(0)}
+\Omega_c^\ast(\sigma_{11}^{(0)}-\sigma_{33}^{(0)})}{|\Omega_c|^2-(\omega
+d_{21})(\omega+d_{31})}F\,e^{i\theta}\nonumber\\
&&=a_{21}^{(1)}F\,e^{i\theta},\nonumber\\
\sigma_{31}^{(1)}&&=\frac{(\omega+d_{21})(\sigma_{11}^{(0)}
-\sigma_{33}^{(0)})+\Omega_c\sigma_{32}^{*(0)}}{|\Omega_c|^2
-(\omega+d_{21})(\omega+d_{31})}F\,e^{i\theta}\nonumber\\
&&=a_{31}^{(1)}F\,e^{i\theta},\label{1st}
\end{eqnarray}
where $F$ is a constant, $\theta=K(\omega) z-\omega t$. The linear dispersion relation
$K(\omega)$ \cite{note2} is given by
\begin{eqnarray}
K(\omega)=\frac{\omega}{c}+\kappa_{13}\int_{-\infty}^{\infty}dv
f(v)\frac{(\omega+d_{21})(\sigma_{11}^{(0)}-\sigma_{33}^{(0)})
+\Omega_c\sigma^{*(0)}_{32}}{|\Omega_c|^2-(\omega+d_{21})(\omega+d_{31})}.\label{eq:LD}
\end{eqnarray}

In thermal equilibrium, $f(v)$ is the Maxwellian velocity
distribution function, i.e.  $f(v)=1/(\sqrt{\pi}\,v_T)\exp
\left[-v^2/v_T^2\right]$, with $v_T=\sqrt{2k_BT/M}$ the most
probable speed at temperature $T$, and $M$ the molecular mass. The
integration in equation (\ref{eq:LD}) with the Maxwellian
distribution leads however to some complicated combination of error
functions~\cite{Ahmed2007}, which is very inconvenient for a simple
and clear analytical approach. As did by Lee {\it et al}.
\cite{Lee2003}, in the following we use the modified Lorentzian
velocity distribution $f(v)=v_T/[\sqrt{\pi}(v_T^2+v^2)]$ to replace
the Maxwellian distribution.

We are interested in two different cases: co-propagating configuration
($k_p\approx k_c$) and counter-propagating configuration
($k_p\approx-k_c$), discussed below separately.

\subsection{Hot molecules with co-propagating configuration}

\begin{figure}
\centering
\includegraphics[scale=0.35]{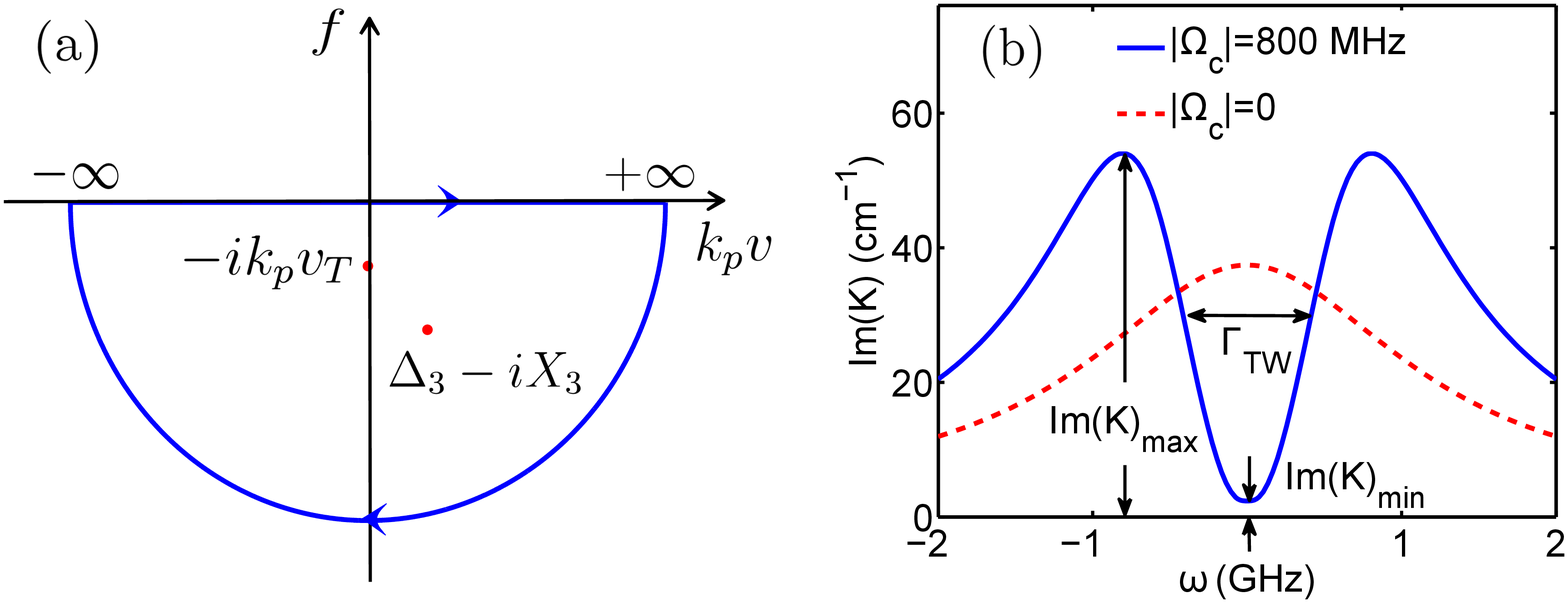}
\caption{(Color online) (a): Two poles $(\Delta_2 ,-iX_3)$,
$(0,-ik_pv_T)$ of the integrand in equation (\ref{eq:LD}) in the
lower half complex plane. The closed curve with arrows is the
contour chosen for calculating the integration in equation
(\ref{eq:LD}) by using residue theorem. (b): Absorption spectrum
Im($K$) as a function of $\omega$ for the hot molecular system. The
solid (dashed) line for $|\Omega_c|=800$ MHz ($|\Omega_c|=0$).
Definitions of Im$(K)_{\rm min}$, Im$(K)_{\rm max}$, and the width
of transparency window $\Gamma_{\rm TW}$ are indicated in the
figure.}\label{fig:LM}
\end{figure}

In this configuration, one has
$d_{21}=\Delta_2-\Delta_1+i\gamma_{21}$,
$d_{31}=-k_pv+\Delta_3-\Delta_1+i\gamma_{31}$ and
$d_{32}=-k_pv+\Delta_3-\Delta_2+i\gamma_{32}$. The second term on
the right-hand side of equation~(\ref{eq:LD}) can be calculated by
using residue theorem~\cite{byr}. There are two poles in the lower
half complex plane
\begin{equation}\label{poles}
\ k_pv=\Delta_3-iX_3, \ \ k_pv=-ik_pv_T,
\end{equation}
with $
X_3\equiv\{\gamma_{32}^2+2\gamma_{32}(2\gamma+\Gamma_{13}+\Gamma_{43})
|\Omega_c|^2/[\gamma(\gamma+\Gamma_3)]\}^{1/2}$. By taking a contour
consisting of real axis and a semi-circle in the lower half complex
plane [see the curves with arrows shown in figure~\ref{fig:LM}(a)],
we can calculate the integration in equation (\ref{eq:LD})
analytically by just calculating the residues corresponding to the
two poles, and obtain exact result for the integration. Since the
expression is lengthy, we just write down the one with
$\Delta_2=\Delta_3=0$, $\Delta\omega_D\gg\gamma_{jl},\ \gamma$:
\begin{eqnarray}\label{CoK}
& & K_{\ } =\frac{\omega}{c}+{\cal K}_1+{\cal K}_2,\\
& &
\mathcal{K}_1=\frac{\sqrt{\pi}\kappa_{13}\Delta\omega_D[2\gamma_{32}(\omega
+i\gamma_{21})A(-iX_3)-iX_3B]}{\gamma\Gamma_{3\gamma}(\Delta\omega_D^2-X_3^2)X_3[|\Omega_c|^2-(\omega
+i\gamma_{21})(\omega+iX_3)]},\nonumber\\
& & \mathcal{K}_2=\frac{\sqrt{\pi}\kappa_{13}[2\gamma_{32}(\omega
+i\gamma_{21})A(-i\Delta\omega_D)-i\Delta\omega_DB]}{\gamma\Gamma_{3\gamma}(X_3^2
-\Delta\omega_D^2)[|\Omega_c|^2
-(\omega+i\gamma_{21})(\omega+i\Delta\omega_D)]},\nonumber
\end{eqnarray}
where $\Delta\omega_D=k_pv_T$ (Doppler width), $A(k_pv)\equiv
X_2\sigma_{11}^{(0)}-\gamma|\Omega_c|^2\sigma_{22}^{{\rm eq}}$ and
$B\equiv\gamma\Gamma_{3\gamma}|\Omega_c|^2\sigma_{22}^{{\rm eq}}$.
Note that $\mathcal{K}_1$ ($\mathcal{K}_2$) is contributed by the
first (second) pole. For cold molecules the second pole in equation
(\ref{poles}) does not exist, thus ${\cal K}_2=0$. However, for hot
molecules one has ${\cal K}_2\neq 0$ due to Doppler effect, and
hence the system may have very different quantum interference
characters comparing with that of cold molecules.

In most cases, $K(\omega)$ can be Taylor expanded around the center
frequency of the probe field (corresponding to $\omega=0$), i.e.,
$K(\omega) =K_0+K_1\omega+(1/2)K_2\omega^2\dots$, where $K_j \equiv
(\partial^jK/\partial\omega^j)_{\omega=0}$. The coefficients $K_0$
describes the phase shift (real part) and the absorption (imaginary
part) per unit length and $1/{\rm Re}(K_1)$ and $1/{\rm Re}(K_2)$
represent the group velocity $v_g$ and group-velocity dispersion,
respectively.

\subsubsection{Transparency window of probe-field absorption spectrum.}

Shown in figure~\ref{fig:LM}(b) is Im($K$) as a function of
$\omega$. The dashed (solid) line is for $|\Omega_c|=0$
($|\Omega_c|=800$ MHz). System parameters given by
$\Gamma_{13}=\Gamma_{23}=\Gamma_{43}=1.77\times10^{7}$
$\mathrm{s}^{-1}$, $\gamma=0.47\times10^{6}$ $\mathrm{s}^{-1}$,
$\gamma_{jl}^{{\rm col}}=4\times10^{6}$ $\mathrm{s}^{-1}$,
$\Delta\omega_D=1.22$ GHz, $\kappa_{13}=5\times10^{10}\ {\rm
cm}^{-1}{\rm s}^{-1}$ and $\sigma_{11}^{{\rm eq}}=\sigma_{22}^{{\rm
eq}}=0.5$. One sees that the absorption spectrum of the probe field
for $|\Omega_c|=0$ has only a single absorption peak. However, a
transparency window opens for a $|\Omega_c=800$ MHz. The minimum
(Im$(K)_{\rm min}$), maximum (Im$(K)_{\rm max}$),  and width of
transparency window ($\Gamma_{\rm TW}$) are defined in the figure.

From equation (\ref{CoK}), we obtain the minimum of Im($K$) at
$\omega=0$:
\begin{equation}\label{eq:A}
{\rm Im}(K)_{\rm min} \simeq \frac{\sqrt{\pi}\kappa_{13}}{\Delta\omega_D}
\left(\frac{\sigma_{11}^{{\rm
eq}}}{1+x_1}-\frac{\sigma_{22}^{{\rm
eq}}}{1+x_1}\frac{1}{1+\sqrt{x}}\right),
\end{equation}
where $x \equiv |\Omega_c|^2\gamma_{31}/(\gamma\Delta\omega_D^2)$
and $x_1\equiv |\Omega_c|^2/(\gamma_{21}\Delta\omega_D)$ are two
dimensionless parameters.  It is interesting that the system has
absorption and gain, reflected by the first and the second terms on
the right hand side of equation (\ref{eq:A}). The gain is due to
non-vanishing $\gamma$ and $\sigma_{22}^{{\rm eq}}$. Obviously, if
$x\gg1$ and $x_1\gg1$, i.e. $|\Omega_c|^2\gamma_{31} \gg
\gamma\Delta\omega_D^2$ and $|\Omega_c|^2 \gg
\gamma_{21}\Delta\omega_D$, one has ${\rm Im}(K)_{\rm min} \approx
0$, i.e. a large and deep transparency widow in the absorption
spectrum is opened.  The inequalities can be taken as the EIT
condition \cite{Li10,Lee2003} of the system. When
$\gamma_{21}\approx\gamma$, this condition is simplified to
$|\Omega_c|^2\gamma_{31}\gg\gamma \Delta\omega_D^2$.

Under the above condition, we obtain ${\rm
Im}(K)_{{\rm max}}\simeq\kappa_{13}\sigma_{11}^{{\rm
eq}}\sqrt{\pi}/\Delta\omega_D$ located at
$\omega\approx\pm\Omega_c$, and
\begin{equation}\label{Tr}
\Gamma_{{\rm TW}}\approx2\left[\frac{2|\Omega_c|^2
+\Delta\omega_D^2-\Delta\omega_D\sqrt{\Delta\omega_D^2+4|\Omega_c|^2}}{2}\right]^{1/2}.
\end{equation}

\subsubsection{EIT-ATS crossover.}

One of our main purposes is to explicitly analyze the detailed
characters of quantum interference effect of the system, which can
be done by extending the spectrum decomposition method introduced
in~\cite{Agarwal1997,Anisimov2008,Tony2010,Anisimov2011}. Note that
${\cal K}_j$ $(j=1,2)$ in equation~(\ref{CoK}) can be decomposed as
\begin{equation}\label{decom1}
{\cal K}_j=\eta_j \left( \frac{A_{j+}}{\omega-\delta_{j+}}+\frac{A_{j-}}{\omega-\delta_{j-}}
\right),
\end{equation}
where $\eta_j$, $A_{j\pm}$ are constants, $\delta_{j+}$ and
$\delta_{j-}$ are two spectrum poles, all of which have been given
explicitly in \ref{App:1}. From equation~(\ref{decom1}) we can get
explicit expressions of Im$({\cal K}_j)$ ($j=1,2$). However, their
general expressions are lengthy and complicated. In order to
illustrate the quantum interference effect in a simple and clear
way, we decompose Im$({\cal K}_j)$ according to different regions of
$\Omega_c$.

(i). {\it Weak control field region}  (i.e. $|\Omega_c|<\Omega_{\rm
ref}\equiv \Delta\omega_D/2$):  In this region, one has
Re$(\delta_{j\pm})$=0, Im$(A_{j\pm})$=0, we obtain
\begin{eqnarray}\label{A_weak}\nonumber
{\rm Im}(K)=\sum_{j=1}^{2}{\rm Im}({\cal
K}_{j})&&=\sum_{j=1}^{2}\eta_{j}\left(\frac{C_{j+}}{\omega^2+W_{j+}^2}
+\frac{C_{j-}}{\omega^2+W_{j-}^2}\right)\\
&&=L_1+L_2,
\end{eqnarray}
where $L_1$ and $L_2$ are defined by
\begin{eqnarray}
&&L_1=\frac{\eta_1C_{1-}}{\omega^2+W_{1-}^2}+\frac{\eta_2C_{2-}}{\omega^2+W_{2-}^2},\nonumber\\
&&L_2=\frac{\eta_1C_{1+}}{\omega^2+W_{1+}^2}+\frac{\eta_2C_{2+}}{\omega^2+W_{2+}^2},\label{L}
\end{eqnarray}
with real constants
\begin{eqnarray}
& & C_{j+}=-W_{j+}(W_{j+}+\Gamma^w_j)/(W_{j+}-W_{j-}),\nonumber\\
& & C_{j-}=W_{j-}(W_{j-}+\Gamma^w_j)/(W_{j+}-W_{j-}),\nonumber\\
& & W_{1\pm}=\frac{1}{2}\left[X_3+\gamma_{21}\pm\sqrt{(X_3-\gamma_{21})^2-4|\Omega_c|^2}\right],\nonumber\\
& & W_{2\pm}=\frac{1}{2}\left[\Delta\omega_D
+\gamma_{21}\pm\sqrt{(\Delta\omega_D-\gamma_{21})^2-4|\Omega_c|^2}\right],\nonumber\\
& &\Gamma_1^w=\gamma_{21}-\frac{X_3B}{2\gamma_{32}A(-iX_3)},\nonumber\\
&
&\Gamma_2^w=\gamma_{21}-\frac{\Delta\omega_DB}{2\gamma_{32}A(-i\Delta\omega_D)}.\label{Wpm}
\end{eqnarray}

\begin{figure}
\centering
\includegraphics[scale=0.3]{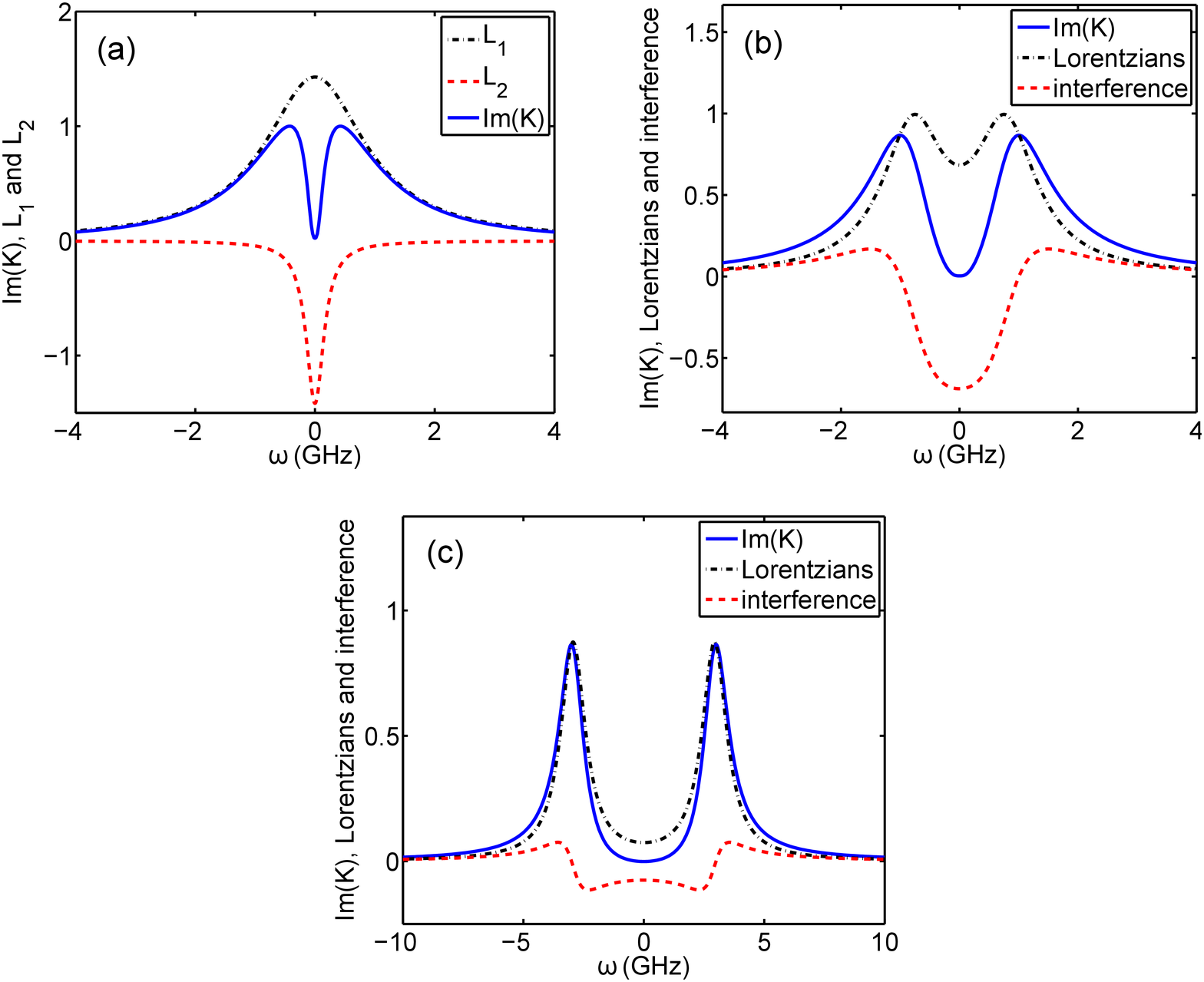}
\caption{(Color online) EIT-ATS crossover for hot molecules in the
co-propagating configuration. (a): Absorption spectrum in the region
$|\Omega_c|<\Omega_{\rm ref}\equiv \Delta\omega_D/2$ contributed by
positive $L_1$ (dashed-dotted line), negative $L_2$ (dashed line),
and total absorption spectrum Im($K$) (solid line). (b) and (c):
Absorption spectrum by two Lorentzians (dashed-dotted line),
destructive interference (dashed line), and total absorption
spectrum Im($K$) (solid line), in the region $|\Omega_c|>\Omega_{\rm
ref}$ and $|\Omega_c|\gg\Omega_{\rm ref}$, respectively. Panels (a),
(b) and (c) correspond to EIT, EIT-ATS crossover, and ATS,
respectively.}\label{co}
\end{figure}

Shown in figure~\ref{co}(a)
are results of $L_1$, which is a positive single peak (the
dashed-dotted line), and $L_2$, which is a negative single peak (the
dashed line). System parameters are given by
$\Gamma_{13}=\Gamma_{23}=\Gamma_{43}=1.77\times10^{7}$
$\mathrm{s}^{-1}$, $\gamma=0.47\times10^{6}$ $\mathrm{s}^{-1}$,
$\gamma_{jl}^{{\rm col}}=4\times10^{6}$ $\mathrm{s}^{-1}$,
$\Delta\omega_D=1.22$ GHz, and $\Omega_c=414$ MHz. The sum of the
positive $L_1$ and negative $L_2$ gives Im(${K}$) (the solid line),
which displays a absorption doublet with a significant transparency
window near at $\omega=0$. Because there exists a {\it destructive}
interference in the probe-field absorption spectrum, the phenomenon
found here belongs to EIT according to the criterion given
in~\cite{Anisimov2008,Tony2010,Anisimov2011}.

(ii). {\it Intermediate control field region} (i.e.
$|\Omega_c|>\Omega_{\rm ref}$): By extending the approach by
Agarwal~\cite{Agarwal1997}, we can decompose Im$({\cal K}_j)$ ($j=1,2$) as
\begin{eqnarray}\label{A_inter}
{\rm Im}&&({\cal
K}_{j})=\eta_{j}\left\{\frac{1}{2}\left[\frac{W_{j}}{(\omega-\delta_{j}^r)^2
+W_{j}^2}+\frac{W_{j}}{(\omega+\delta_{j}^r)^2+W_{j}^2}\right]\right.\nonumber\\
&&\hspace{0.3cm}\left.+\frac{g_{j}}{2\delta_{j}^r}\left[\frac{\omega-\delta_{j}^r}{(\omega
-\delta_{j}^r)^2+W_{j}^2}-\frac{\omega+\delta_{j}^r}{(\omega+\delta_{j}^r)^2+W_{j}^2}\right]\right\},
\end{eqnarray}
where
\begin{eqnarray}
&&W_1=(\gamma_{21}+X_3)/2,\nonumber\\
&&W_2=(\gamma_{21}+\Delta\omega_D)/2,\nonumber\\
&&\delta^r_{1}=\sqrt{4|\Omega_c|^2-(X_3-\gamma_{21})^2}/2,\nonumber\\
&&\delta^r_{2}=\sqrt{4|\Omega_c|^2-(\Delta\omega_D-\gamma_{21})^2}/2,\nonumber\\
&&g_1=\frac{X_3-\gamma_{21}}{2}+\frac{X_3B}{2\gamma_{32}A(-iX_3)},\nonumber\\
&&g_2=\frac{\Delta\omega_D-\gamma_{21}}{2}+\frac{\Delta\omega_DB}{2\gamma_{32}A(-i\Delta\omega_D)}.\label{CS1co}
\end{eqnarray}
The first two terms in the first square bracket on the right hand
side of equation~(\ref{A_inter}) are two Lorentzians, resulted from
the absorption from two different pathways corresponding to the two
dressed states created by the coupling field. The terms in the
second square bracket are interference terms, the magnitudes of
which are controlled by the parameter $g_j$. If $g_j>0$ ($g_j<0$)
the interference is destructive (constructive).

Figure~\ref{co}(b) shows the result of the probe-field absorption
spectrum as functions of $\omega$ for $|\Omega_c|>\Omega_{\rm ref}$.
The dashed-dotted line (dashed line) denotes the contribution by two
Lorentzians  (interference terms). We see that the interference is
destructive. The solid line gives the result of Im($K$). System
parameters used are the same as those in panel (a) but with
$\Omega_c=1$ GHz. A transparency window opens due to the combined
effect of EIT and ATS, which is deeper and wider than that in panel
(a). We call such phenomenon as EIT-ATS crossover.

(iii). {\it Large control field region} (i.e.
$|\Omega_c|\gg\Omega_{\rm ref}$): In this case, the quantum
interference strength $g_j/\delta_j^r$ in equation~(\ref{A_inter})
is very weak and negligible. We have
\begin{equation}\label{A_strong}
{\rm Im}({\cal K}_j)\approx
\frac{\eta_j}{2}\left[\frac{W_j}{(\omega-\delta_j^r)^2+W_j^2}
+\frac{W_j}{(\omega+\delta_j^r)^2+W_j^2}\right],
\end{equation}
being to a sum of two Lorentzians.

Shown in the panel (c) of figure~\ref{co} is the result of the
probe-field absorption spectrum as functions of $\omega$ for
$|\Omega_c| \gg \Omega_{\rm ref}$. The dashed-dotted line represents
the contribution by the sum of the two Lorentzians. For
illustration, we have also plotted the contribution from the small
interference terms [neglected in equation (\ref{A_strong})\,],
denoted by the dashed line. We see that the interference is still
destructive but very small. The solid line is the curve of Im($K$),
which has two resonances at $\omega\approx\pm\Omega_c$. Parameters
used are the same as those in panel (a) and (b)  but with
$\Omega_c=3$ GHz. Obviously, the phenomenon found in this situation
belongs to ATS because the transparency window opened is mainly due
to the contribution of the two Lorenztians.

\begin{figure}
\centering
\includegraphics[scale=0.4]{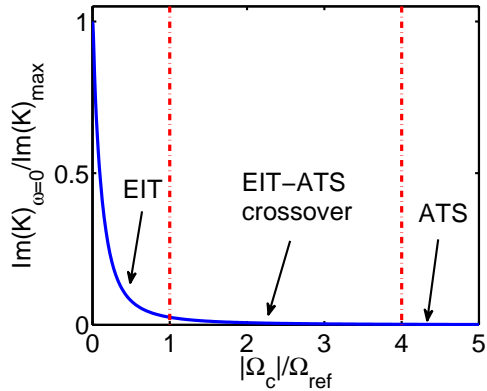}\\
\caption{(Color online) The ``phase diagram'' illustrating the
transition from EIT to ATS for hot molecules in the co-propagating
configuration. Shown is ${\rm Im}(K)_{\omega=0}/{\rm Im}(K)_{\rm
max}$ as a function of $|\Omega_c|/\Omega_{\rm ref}$. Three regions
(EIT, EIT-ATS crossover, and ATS) are divided by two dash-dotted
lines.}\label{ratio_co}
\end{figure}

From above results, we see that the probe-field absorption spectrum
experiences a transition from EIT to ATS as $\Omega_c$ is changed
from weak to strong values. Since in three-level systems such
phenomenon happens quite often and is universal, we divide quantum
interference effects into three classes, i.e. the EIT region
($|\Omega_c|<\Omega_{\rm ref}$), the region of the EIT-ATS crossover
($1< |\Omega_c|/\Omega_{\rm ref}\leq 4$), and ATS region
$(|\Omega_c|/\Omega_{\rm ref}> 4$). Figure~\ref{ratio_co} shows a
``phase diagram'' that
illustrates the transition from the EIT to ATS  by plotting ${\rm
Im}(K)_{\omega=0}/{\rm Im}(K)_{\rm max}$ as a function of
$|\Omega_c|/\Omega_{\rm ref}$. Note that we have defined
${\rm Im}(K)_{\omega=0}/{\rm Im}(K)_{\rm
max}=0.01$ as the border between EIT-ATS crossover and ATS regions.

\subsubsection{Comparison with experiment.}

\begin{figure}
\centering
\includegraphics[scale=0.45]{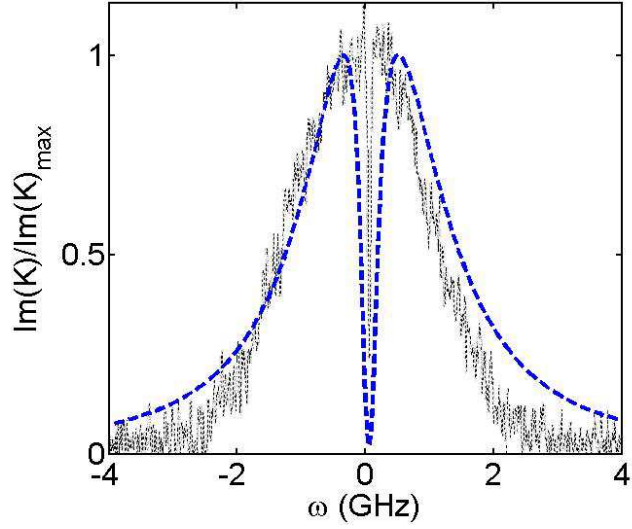}
\caption{(Color online) Probe-field absorption spectrum ${\rm
Im}(K)/{\rm Im}(K)_{{\rm max}}$ as a function of frequency $\omega$,
with $\Omega_c=414\ {\rm MHz}$ (EIT region). The dashed line is
theoretical result. The solid line is the experimental one
reported in Ref.~\cite{Lazoudis2010}.}\label{exp}
\end{figure}

To check the theoretical prediction given above, it is necessary to
make a comparison with the experiment reported recently by Lazoudis
{\it et al.}~\cite{Lazoudis2010}, which was performed with a
co-propagating configuration. Using system parameters
$\Gamma_{13}=\Gamma_{23}=\Gamma_{43}=1.77\times10^{7}$
$\mathrm{s}^{-1}$, $\gamma=0.47$ MHz, $\gamma_{jl}^{{\rm col}}=4$
MHz, and $\Delta\omega_D=1.22$ GHz, we have calculated probe-field
absorption spectrum ${\rm Im}(K)$ as a function of frequency
$\omega$, with $\Omega_c=414\ {\rm MHz}$ (EIT region) and the
control-field detuning $-55$ MHz. The result is plotted as the
dashed line of figure~\ref{exp},
%
%
which agrees fairly with the experimental one (the solid line)
measured in~\cite{Lazoudis2010} (see figure~5(a)
of~\cite{Lazoudis2010}). Note that here we have plotted the quantity
${\rm Im}(K)$, which is proportional to fluorescence intensity
(measured in~\cite{Lazoudis2010}) related to the state $|3\rangle$
because $\sigma_{33}\simeq2|\Omega_p|^2{\rm
Im}(K)/(\gamma+\Gamma_3)$~\cite{note3}. The small difference for
depth and width of the EIT dip between  our result and the
experiment is due to the approximation by using the modified
Lorentzian distribution to replace the Maxwellian velocity
distribution.

\subsection{Hot molecules with counter-propagating configuration}

We now move to the situation when the probe  and
control fields are arranged as a counter-propagating configuration.
Here, $d_{21}=\Delta_2-\Delta_1-2k_pv+i\gamma_{21}$ and
$d_{32}=\Delta_3-\Delta_2+k_pv+i\gamma_{32}$. Then we obtain
\begin{eqnarray}
& & K_{\ }=\frac{\omega}{c}+\frac{\kappa_{13}}{\gamma\Gamma_{3\gamma}}(\mathcal{K}_1+\mathcal{K}_2),\label{CounterK}\\
& &
\mathcal{K}_1=\frac{\sqrt{\pi}\Delta\omega_D[2\gamma_{32}(\omega+i2X_3)
A(-iX_3)+iX_3B]}{(\Delta\omega_D^2-X_3^2)X_3
[|\Omega_c|^2-(\omega+2iX_3)(\omega+iX_3)]},\nonumber\\
& &
\mathcal{K}_2=\frac{\sqrt{\pi}[2\gamma_{32}(\omega+i2\Delta\omega_D)
A(-i\Delta\omega_D)+i\Delta\omega_DB]}{(X_3^2-\Delta\omega_D^2)
[|\Omega_c|^2-(\omega+i2\Delta\omega_D)
(\omega+i\Delta\omega_D)]},\nonumber
\end{eqnarray}
where ${\cal K}_1$ and ${\cal K}_2$ are obtained from the poles
$k_pv=\Delta_3-iX_3$ and $k_pv=-ik_pv_T$, respectively.

\begin{figure}
\centering
\includegraphics[scale=0.3]{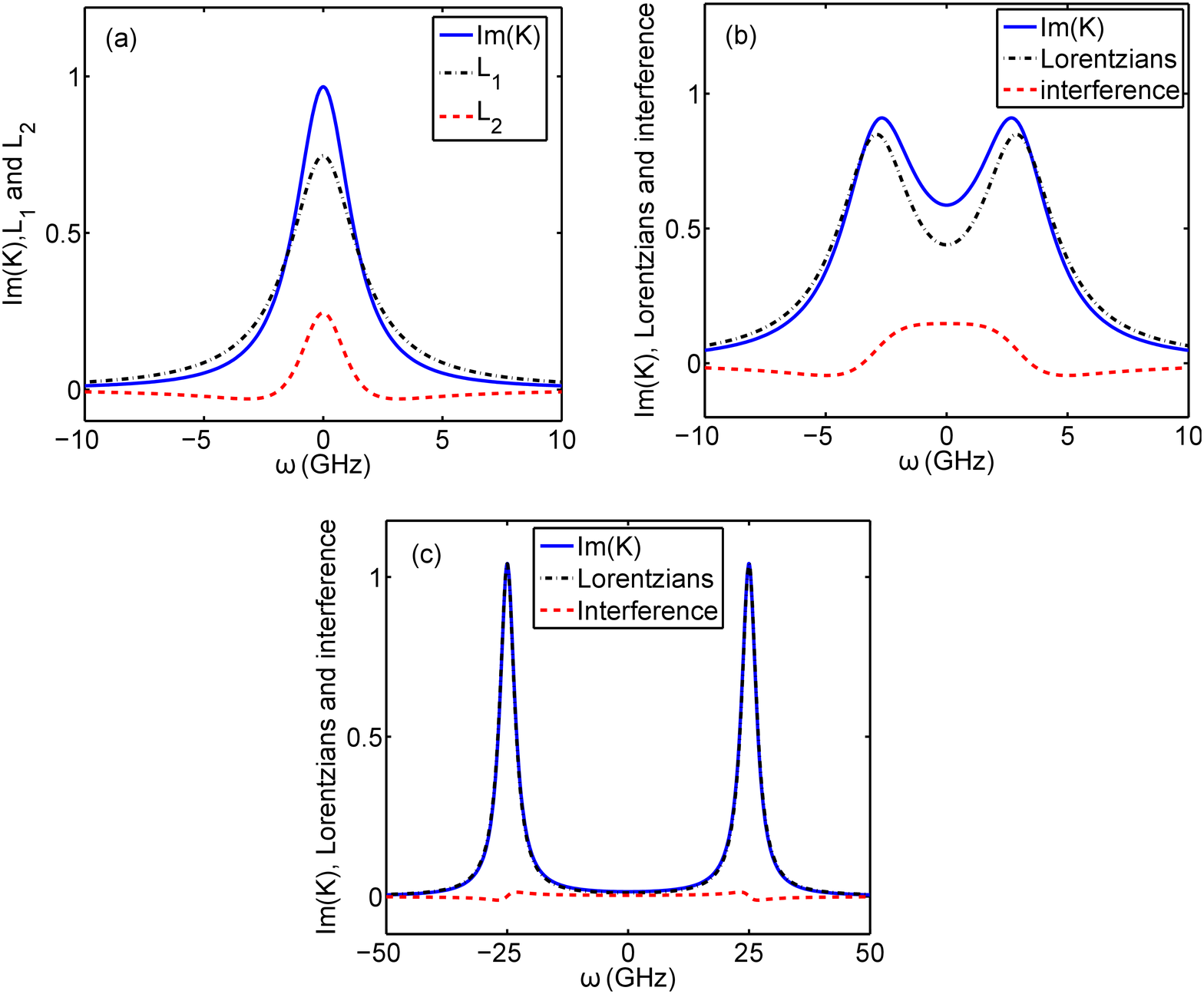}
\caption{(Color online) Probe-field absorption spectrum for hot
molecules in the counter-propagating configuration. (a): Absorption
spectrum in the region $|\Omega_c|<\Omega_{\rm ref}\equiv
\Delta\omega_D/2$ contributed by positive $L_1$ (dashed-dotted line)
and $L_2$ (dashed line), and total absorption spectrum Im($K$)
(solid line). (b) and (c): Absorption spectrum by two Lorentzians
(dashed-dotted line), constructive interference (dashed line), and
total absorption spectrum Im($K$) (solid line), in the region
$|\Omega_c|>\Omega_{\rm ref}$ and $|\Omega_c|\gg\Omega_{\rm ref}$,
respectively.}\label{counter}
\end{figure}

We have carried out a similar spectrum decomposition as that did for
the co-propagating configuration given above. For saving space, here
we omit concrete expressions of the spectrum decomposition but
present probe-field absorption spectra in three typical
control-field regions in figure~\ref{counter}.

Shown in the panel (a) of figure~\ref{counter}
is the result of probe-field absorption spectrum Im($K$) in weak
control-field region (i.e. $|\Omega_c|<\Omega_{\rm ref}$) as a
function of $\omega$ for $\Omega_c=500$ MHz. As in
figure~\ref{co}(a), Im($K$) is also the sum of two terms, i.e. $L_1$
and $L_2$. Nevertheless, now both $L_1$ and $L_2$ are positive, as
illustrated by the dashed-dotted line and dashed line, respectively.
We see that Im($K$) (the solid line) displays only a positive single
peak, there is no transparency window, and the reason is that the
quantum interference becomes constructive (the red dashed line) for
the counter-propagating configuration. Thus, different from the case
of the co-propagating configuration, in weak control-field region an
EIT which we have defined as transparency window plus a destructive
interfrence does not exists.

Shown in figure~\ref{counter}(b) and (c) are results of the
probe-field absorption spectra as functions of $\omega$ for
$|\Omega_c|>\Omega_{\rm ref}$ and $|\Omega_c|\gg\Omega_{\rm ref}$,
respectively. System parameters are given by
$\Gamma_{13}=\Gamma_{23}=\Gamma_{43}=1.77\times10^{7}$
$\mathrm{s}^{-1}$, $\gamma=0.47\times10^{6}$ $\mathrm{s}^{-1}$,
$\gamma_{ij}^{{\rm col}}=4\times10^{6}$ $\mathrm{s}^{-1}$, and
$\Delta\omega_D=1.22$ GHz, with $\Omega_c=3$ GHz (in the
intermediate control-field region) and $\Omega_c=25$ GHz (in the
large control-field region)  for the panel (b) and the panel (c),
respectively. The dashed-dotted line (dashed line) denotes the
contribution by the sum of two Lorentzians terms (interference
terms) in Im($K$). The solid line gives the result of Im($K$). We
see that the interferences near the probe-field center frequency
(i.e. $\omega=0$) are always constructive. Consequently, different
from the case of the co-propagating configuration, no EIT-ATS
crossover happens.

\begin{figure}
\centering
\includegraphics[scale=0.35]{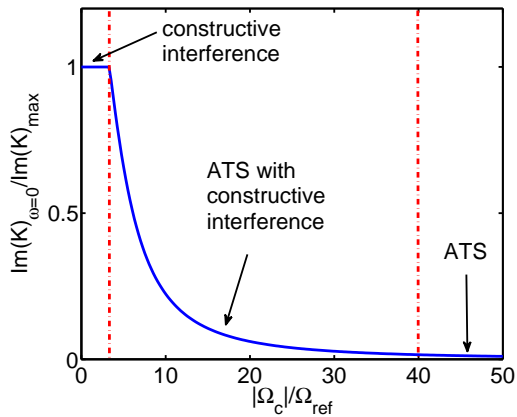}
\caption{(Color online) ${\rm Im}(K)_{\omega=0}/{\rm Im}(K)_{\rm
max}$ as the function of the control field $|\Omega_c|/\Omega_{\rm
ref}$ for hot molecules in the counter-propagating configuration.
Three regions (constructive interference, ATS with constructive
interference and ATS) are divided by two dashed-dotted
lines.}\label{ratio_counter}
\end{figure}

Shown in figure~\ref{ratio_counter}
is the  ``phase diagram'' that illustrates the transition from the
constructive interference to ATS  for the counter-propagating
configuration by plotting ${\rm Im}(K)_{\omega=0}/{\rm Im}(K)_{\rm
max}$ as a function of $|\Omega_c|/\Omega_{\rm ref}$. Three regions
are divided as constructive interference, ATS with constructive
interference, and ATS, respectively.

\subsection{Cold molecules and comparison for various cases }\label{sec_cold}

Our model presented in section~\ref{Sec:2} is also valid for cold
molecules. In this case, one should take $v=0$ in the Bloch equation
(\ref{dme}),  and $f(v)=\delta (v)$ in the Maxwell
equation (\ref{eqs:maxwell}). The solutions (\ref{BS}) and
(\ref{1st}) are still valid but one must take $v=0$ there. However,
the dispersion relation (\ref{eq:LD}) is replaced by
\begin{equation}\label{LD_cold}
K(\omega)=\frac{\omega}{c}+\frac{\kappa_{13}(\sigma_{11}^{(0)}
-\sigma_{33}^{(0)})(\omega+i\Gamma)}{|\Omega_c|^2
-(\omega+i\gamma_{21})(\omega+i\gamma_{31})},
\end{equation}
with
$\Gamma=\gamma_{21}+|\Omega_c|^2(\sigma_{33}^{(0)}
-\sigma_{22}^{(0)})/[\gamma_{32}(\sigma_{11}^{(0)}-\sigma_{33}^{(0)})]$.
Here $\Delta_2=\Delta_3=0$ has been taken for simplicity.

\begin{figure}
\centering
\includegraphics[scale=0.4]{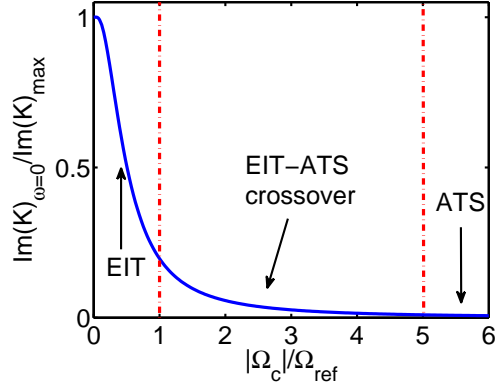}
\caption{(Color online) Transition from EIT to ATS for cold
molecules. Shown is ${\rm Im}(K)_{\omega=0}/{\rm Im}(K)_{\rm max}$
as a function of $|\Omega_c|/\Omega_{\rm ref}$, where $\Omega_{\rm
ref}\equiv |\gamma_{21}-\gamma_{31}|/2$. Three regions (EIT, EIT-ATS
crossover and ATS) are divided by two dash-dotted
lines.}\label{ratio_cold}
\end{figure}

A similar spectrum decomposition can be done like that did for hot
molecules, which is omitted here.  Shown in figure~\ref{ratio_cold}
is the probe-field absorption spectrum ${\rm Im}(K)_{\omega=0}/{\rm
Im}(K)_{\rm max}$ as a function of the control field
$|\Omega_c|/\Omega_{\rm ref}$, where $\Omega_{\rm ref}\equiv
|\gamma_{21}-\gamma_{31}|/2$. System parameters are given by
$\Gamma_{13}=\Gamma_{23}=\Gamma_{43}=1.77\times10^{7}$
$\mathrm{s}^{-1}$, $\gamma_{ij}^{{\rm col}}=1\times10^{3}$
$\mathrm{s}^{-1}$ and $\sigma_{11}^{\rm eq}=1$. From the figure, we
obtain the similar conclusion as that obtained for co-propagation
configuration, i.e. the probe-field absorption
spectrum experiences also a transition from EIT to ATS as
$\Omega_c$ is increased. The quantum interference effect in the system
can divided into three regions, i.e. the EIT region ($|\Omega_c|<\Omega_{\rm
ref}$), the region of the EIT-ATS crossover ($1< |\Omega_c|/\Omega_{\rm ref}\leq 5$),
and ATS region $(|\Omega_c|/\Omega_{\rm ref}> 5$).

From the results given above, we see that
the quantum coherence in the open $\Lambda$-type molecular system
has very interesting features, depending on the existence or
non-existence of the Doppler broadening, and also depending on the
beam propagating (co-propagating or counter-propagating)
configurations. For comparison, in Table~\ref{table:EIT} some useful
physical quantities, including EIT condition, absorption spectrum ${\rm
Im}(K)|_{\omega=0}$, group velocity $v_{g}$, and width of
transparency window $\Gamma_{{\rm TW}}$, are presented for several
different physical cases.
%
\begin{table}
\caption{Propagating properties of the probe field for
various open $\Lambda$-type molecular systems, including EIT condition,
absorption spectrum Im$(K)|_{\omega=0}$, width of transparency window
$\Gamma_{\rm TW}$, and group velocity $v_g$ for three different
cases. Other quantities appeared in the Table have been defined in
the text. Mol.=Molecules, Co-prop.=Co-propagating configuration,
Cou.-prop.=Counter-propagating configuration.\label{table:EIT}}
\begin{tabular}{lllll}
\br
System & EIT condition &  Im$(K)|_{\omega=0}$& $\Gamma_{\rm TW}$ & $v_g$ \\
\mr
Hot Mol. (Co-prop.)& $\frac{\gamma\Delta\omega_D^2}{\gamma_{31}}\leq |\Omega_c|^2\leq\frac{(\Delta\omega_D)^2}{4}$ & $\frac{\sqrt{\pi}\kappa_{13}\gamma_{21}}{|\Omega_c|^2}$ & $\frac{2|\Omega_c|^2}{\Delta\omega_D}$ & $\frac{|\Omega_c|^2}{\sqrt{\pi}\kappa_{13}}$\\
Hot Mol. (Cou.-prop.)   & no EIT & $\frac{\sqrt{\pi}\kappa_{13}\Delta\omega_D}{|\Omega_c|^2}$ & $2|\Omega_c|-\Delta\omega_D$ & $\frac{|\Omega_c|^2}{\sqrt{\pi}\kappa_{13}}$\\
Cold Mol.  & $\gamma_{21}\gamma_{31}\leq|\Omega_c|^2\leq \frac{\gamma_{31}^2}{4}$ &$ \frac{\kappa_{13}\gamma_{21}}{|\Omega_c|^2}$ & $\frac{2|\Omega_c|^2}{\gamma_{31}}$ & $\frac{|\Omega_c|^2}{\kappa_{13}}$\\
\br
\end{tabular}
\end{table}

The first line in the Table is for hot molecules working in the
co-propagating configuration; the second line is for hot molecules
working in the counter-propagating configuration; the third line is for cold molecules.
There are EIT, EIT-ATS crossover, and ATS for both cold molecules and the
hot molecules with the co-propagating configuration. But there is no EIT and
no EIT-ATS crossover for the hot molecules with the counter-propagating configuration.
Experimentally, up to now only the EIT in the co-propagating configuration has been
demonstrated recently by experiment~\cite{Lazoudis2010}.

\section{Nonlinear pulse propagation}{\label{sec_soliton}}

The theoretical approach given in the last two sections is valid not only for continuous-wave
but also for pulsed probe fields. However, if the probe field is pulsed and has a larger amplitude,
nonlinear effect induced by Kerr nonlinearity inherent in the system must taken into account.
We stress that the theoretical scheme proposed in the present work is very suitable for the study
of pulse propagation in multi-level systems.

In this section, we investigate nonlinear pulse propagation,
especially ultraslow optical solitons, in the present open hot
molecular system with co-propagating configuration by using the
method of multiple-scales. For this aim, we take the asymptotic
expansion
$\sigma_{jl}-\sigma_{jl}^{(0)}=\sum_{m=1,2,\cdots}\epsilon^{m}\sigma_{jl}^{(m)}$,
$\Omega_{p}=\sum_{m=1,2,\cdots}\epsilon^{m}\Omega_{p}^{(m)}$, with
$\sigma_{jj}^{(1)}=0$ and $\sigma_{32}^{(1)}=0$, where $\epsilon$ is
a small parameter denoting the typical amplitude of $\Omega_{p}$ and
all quantities on the  right hand side of the asymptotic expansion
are considered as functions of the multi-scale variables
$z_m=\epsilon^mz\ (m=0,1,2)$, $t_m=\epsilon^m t$ $(m=0,1)$.
Substituting the expansion into the MB equations (\ref{dme}) and
(\ref{eqs:maxwell}), we obtain a series of linear but
inhomogeneous equations for $\sigma_{ij}^{(m)}$ and $\Omega_p^{(m)}$
($m=1$-4), which can be solved order by order.

The zeroth-order ($m=0$) and the first-order ($m=1$) solutions are
the same as that given respectively by equation~(\ref{BS}) and
(\ref{1st}), by now $\theta=K(\omega) z_0-\omega t_0$ and $F$ is yet
to be determined envelope function of the ``slow'' variables $t_1$,
$z_1$ and $z_2$. In the second order ($m=2$), a divergence-free
solution for $\Omega_p^{(2)}$ requires the solvability condition
$i[\partial F/\partial z_1+(\partial K/\partial\omega)\partial
F/\partial t_1]=0$, which shows that the envelope function $F$
travels with complex group velocity $(\partial
K/\partial\omega)^{-1}$. Explicit expressions of the second order
solution have been given in \ref{App:2}.

In the third order ($m=3$), the Kerr nonlinearity of the system plays a role. A
divergence-free solution for $\Omega_{p}^{(3)}$ gives rise to the equation
\begin{equation}\label{solvability3}
i\frac{\partial F}{\partial
z_{2}}-\frac{1}{2}\frac{\partial^{2}K}{\partial\omega^{2}}\frac{\partial^{2}F}{\partial
t_{1}^{2}}-W|F|^{2}Fe^{-2\bar{\alpha} z_{2}}=0,
\end{equation}
where $\alpha={\rm Im}(K)=\epsilon^2\bar{\alpha}$ and
\begin{equation}\label{W}
W=-\kappa_{13}\int_{-\infty}^{\infty}dv
f(v)\frac{\Omega_{c}a_{32}^{\ast(2)}+(\omega+d_{21})(a_{11}^{(2)}-a_{33}^{(2)})}
{|\Omega_{c}|^{2}-(\omega+d_{21})(\omega+d_{31})},
\end{equation}
with coefficients $a_{11}^{(2)}$, $a_{22}^{(2)}$ and $a_{32}^{(2)}$
are defined in \ref{App:2}.

Combining equation~(\ref{solvability3}) and the solvability
condition in the second order, we obtain
\begin{equation}\label{NLSE1}
i\frac{\partial}{\partial z}U-\frac{1}{2}\frac{\partial^{2}K}
{\partial\omega^{2}}\frac{\partial^{2}
U}{\partial\tau^{2}}-W|U|^{2}Ue^{-2\alpha z}=0,
\end{equation}
where $\tau=t-z/v_{g}$ and $U=\epsilon F$.  Equation (\ref{NLSE1}) is a nonlinear Schr\"{o}dinger (NLS) equation describing time evolution of the envelope function $F$, in which $W$ is proportional to third-order nonlinear susceptibility (Kerr coefficient) relevant to self-phase modulation, which is necessary for the formation of a shape-preserved probe pulse.

The key for the formation and propagation of an optical soliton in
the system requires two conditions. The first is a balance between
dispersion and nonlinearity, and the second is the absorption of the
probe field must be negligibly small.  Generally, the coefficients
of the equation~(\ref{NLSE1}) are complex, which means that a
soliton, even if it is produced initially, may be highly unstable
during propagation. However, as shown below, a realistic set of
system parameters can be found under the EIT condition so that the
imaginary part of these coefficients can be made much smaller than
their corresponding real part. Thus it is possible to get a
shape-preserving nonlinear localized solution that can propagate a
rather long distance without a significant distortion.

Neglecting the small imaginary part of the coefficients and taking
$\omega=0$, equation (\ref{NLSE1}) can be written into the
dimensionless form $i\partial u/\partial s+\partial^{2}u/\partial
\sigma^{2}+2|u|^{2}u =0$, with $s=-z/(2L_{D})$,
$\sigma=\tau/\tau_{0}$, and $u=U/U_{0}$. Here $\tau_0$ is typical
pulse duration, $L_{D}=\tau_{0}^{2}/\widetilde{K}_2$ is typical
dispersion length, and
$U_{0}=(1/\tau_{0})\sqrt{\widetilde{K}_2/\widetilde{W}}$ is typical
half Rabi frequency of the probe field,  with $\widetilde{K}_2$ and
$\widetilde{W}$ being the real part of
$K_2=(\partial^2K/\partial\omega^2)_{\omega=0}$ and $W|_{\omega=0}$,
respectively. Then one can obtain various soliton solutions for $u$.
A single-soliton solution in terms of the half Rabi frequency reads
\begin{equation}\label{SOLITON 2}
\Omega_{p}=\frac{1}{\tau_0}
\sqrt{\frac{\widetilde{K_2}}{\widetilde{W}}} {\rm sech}\,
\left(\frac{t}{\tau_0}-\frac{z}{\tau_0v_g}\right) \,\exp \left[
i\left(\widetilde{K}_0 +\frac{1}{2L_D}\right)z \right]
\end{equation}
with $\widetilde{K}_{0}=\mathrm{Re}(K)|_{\omega=0}$, which describes a bright soliton
traveling with the propagating velocity
$v_g=[{\rm Re}(\partial K/\partial \omega)]^{-1}|_{\omega=0}$.

We now give a realistic parameter set for the formation of the
optical soliton given above. For a hot Li$_2$ molecular gas, we
choose  $\Omega_{c}=600$ $\mathrm{MHz}$,
$\Delta_2=\Delta_{3}\approx2.36\times10^{7}$ $\mathrm{s}^{-1}$,
$\tau_{0}=1.0\times10^{-7}$ $\mathrm{s}$,
$\omega_p=4.46\times10^{14}\ \rm{s}^{-1}$, and other parameters are
the same as those given in the previous text. Then we obtain
$K_{2}=(5.51+0.672i)\times10^{-16}$ $\mathrm{cm}^{-1}\mathrm{s}^{2}$
and $W=(1.75+0.298i)\times10^{-16}$
$\mathrm{cm}^{-1}\mathrm{s}^{2}$, $L_{D}=L_{NL}=18.2$ $\mathrm{cm}$,
and  $U_0=1.77\times10^7\ {\rm s}^{-1}$. One sees that the imaginary
part of $K_2$ and $W$ is indeed much smaller than their
corresponding real part. The reason of so small imaginary part is
due to the quantum interference effect contributed by the control field.

The propagating velocity of the probe pulse can be estimated by the
real part of the linear dispersion relation (\ref{eq:LD}). At the
probe-field center frequency (i.e. $\omega=0$) we obtain  $v_g=[{\rm
Re}(\partial K/\partial \omega)|_{\omega=0}]^{-1}\approx
2.13\times10^{-4}c$. Consequently, the optical soliton obtained may
travel with an ultraslow propagating velocity in the system.

\begin{figure}
\centering
\includegraphics[scale=0.4]{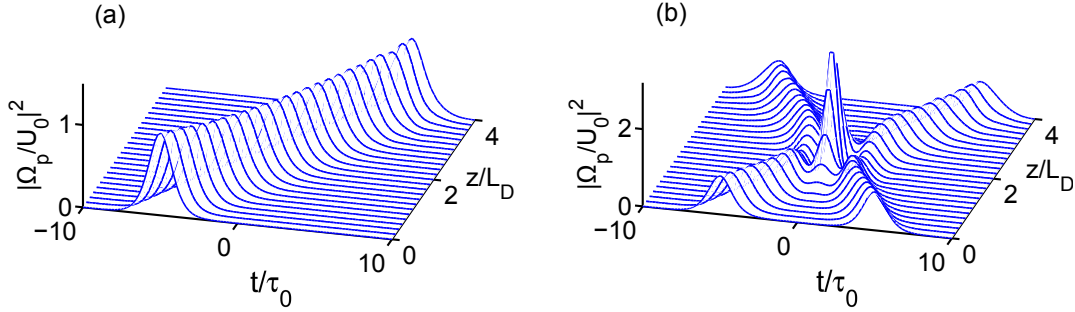}
\caption{(Color online) (a): Ultraslow optical solitons and their
interaction in the hot molecular system. (a): Three-dimensional plot
of the waveshape of $|\Omega_p/U_{0}|^2$ as a function of $z/L_D$
and $t/\tau_0$. (b): Collision between two ultraslow optical
solitons.}\label{soli}
\end{figure}

The stability of the ultraslow optical soliton described above can
be checked by using numerical simulations. In figure~\ref{soli}(a),
we show the wave shape of $|\Omega_p/U_0|^2$ as a function of
$z/L_D$ and $t/\tau_0$. The solution is obtained by numerically
solving Eq.~(\ref{NLSE1}) with full complex coefficients
included. The initial condition is given by $\Omega_p(0,t)=U_0{\rm
sech}(t/\tau_0)$. We see that the amplitude of the soliton undergoes
only a slight decrease and its width undergoes a slight increase due
to the influence of the imaginary part of the coefficients. A
simulation of the interaction between two ultraslow optical solitons
is also carried out by inputting two identical solitons [see
figure~\ref{soli} (b)]. The initial condition is
$\Omega_p(0,t)=U_0{\rm sech}(t/\tau_0-5)+U_0{\rm sech}(t/\tau_0+5)$.
As time goes on, they collide, pass through, and depart from each
other. The two solitons recover their initial waveforms after the
collision. However, a phase shift is observed after the collision.

\section{CONCLUSION}\label{sec_conclusion}

We have developed a systematic analytical approach on linear and nonlinear
pulse propagations in an open $\Lambda$-type molecular system with
Doppler broadening. In linear case, by using residue theorem and
spectrum decomposition method, we have proved that there exists a crossover
from EIT to ATS for the co-propagating configuration. However, there is no EIT
and hence no EIT-ATS crossover for the counter-propagating configuration. We have provided
various explicit formulas, including probe-field spectrum decomposition,
EIT condition, and width of EIT transparency window, as well as a comparison
with the result of cold molecules. Our analytical result agrees well
with the experimental one reported recently by
Lazoudis {\it et al}~\cite{Lazoudis2010}. In nonlinear case, by using the method of multiple-scales,
we have derived a nonlinear envelope equation for probe-field propagation.
We show that stable ultraslow solitons can be realized in the open
molecular system. New theoretical predictions presented in this work are
helpful for guiding new experimental findings in coherent molecular systems
and may have promising practical applications in coherent molecular spectroscopy,
precision measurement, molecular quantum state control, nonlinear pulse
propagation, and so on.

\begin{ack}
This work was supported by NSF-China under Grant Numbers 10874043 and
11174080.
\end{ack}

\appendix

\section{Expressions of $\eta_j$, $A_{j\pm}$, and $\delta_{j\pm}$}\label{App:1}
\label{corpoles}
\begin{eqnarray}
&&\eta_1=\frac{\kappa_{13}\sqrt{\pi}\gamma_{32}\Delta\omega_DA(-iX_3)}{\gamma\Gamma_{3\gamma}X_3(\Delta\omega_D^2-X_3^2)},\\
&&\eta_2=\frac{\kappa_{13}\sqrt{\pi}\gamma_{32}A(-i\Delta\omega_D)}{\gamma\Gamma_{3\gamma}(X_3^2-\Delta\omega_D^2)},\\\nonumber
&&\delta_{1\pm}=\frac{1}{2}\left[-i(X_3+\gamma_{21})\pm\sqrt{4|\Omega_c|^2-(X_3-\gamma_{21})^2}\right],\\
&&\\\nonumber
&&\delta_{2\pm}=\frac{1}{2}\left[-i(\Delta\omega_D+\gamma_{21})\pm\sqrt{4|\Omega_c|^2-(\Delta\omega_D-\gamma_{21})^2}\right],\\
&&\\
&&A_{1\pm}=\mp\frac{\delta_{1\pm}-\left[\gamma_{21}-\frac{X_3B}{2\gamma_{32}A(-iX_3)}\right]}{\delta_{1+}-\delta_{1-}},\\
&&A_{2\pm}=\mp\frac{\delta_{2\pm}-\left[\gamma_{21}-\frac{\Delta\omega_DB}{2\gamma_{32}A(-i\Delta\omega_D)}\right]}{\delta_{2+}-\delta_{2-}}.
\end{eqnarray}

\section{Second-order solution of MB Equations}\label{App:2}

\begin{eqnarray}\nonumber
\sigma_{21}^{(2)}&&=\frac{i}{D}[(\omega+d_{31})a_{21}^{(1)}-\Omega_c^*a_{31}^{(1)}]\frac{\partial
F}{\partial t_{1}}e^{i\theta}\\
&&=a_{21}^{(2)}\frac{\partial F}{\partial
t_{1}}e^{i\theta},\\\nonumber
\sigma_{31}^{(2)}&&=\frac{i}{D}[(\omega+d_{21})a_{31}^{(1)}-\Omega_ca_{21}^{(1)}]\frac{\partial
F}{\partial t_{1}}e^{i\theta}\\
&&=a_{31}^{(2)}\frac{\partial F}{\partial
t_{1}}e^{i\theta},\\\nonumber
\sigma_{33}^{(2)}&&=\frac{i}{D_1}\left\{\left[\gamma^2(\omega+d_{32})(\omega+d_{32}^{*})+2\gamma\gamma_{32}|\Omega_c|^2\right]\right.\\\nonumber
&&\times(a_{31}^{*(1)}-a_{31}^{(1)})-\gamma(\gamma+\Gamma_{31})\left[\Omega_ca_{21}^{(1)}(\omega+d_{32})\right.\\\nonumber
&&\left.\left.-\Omega_c^*a_{21}^{*(1)}(\omega+d_{32}^{*})\right]\right\}|F|^2e^{-2\bar{\alpha}z_2}\nonumber\\
&&=a_{33}^{(2)}|F|^2e^{-2\bar{\alpha}z_2},\\\nonumber
\sigma_{11}^{(2)}&&=\left[\frac{\Gamma_{13}}{\gamma+\Gamma_{31}}a_{33}^{(2)}-\frac{i}{\gamma+\Gamma_{31}}(a_{31}^{*(1)}-a_{31}^{(1)})\right]|F|^2e^{-2\bar{\alpha}z_2}\\
&&=a_{11}^{(2)}|F|^2e^{-2\bar{\alpha}z_2},\\
\sigma_{22}^{(2)}&&=-(\sigma_{11}^{(2)}+\sigma_{33}^{(2)}+\sigma_{44}^{(2)})=a_{22}^{(2)}|F|^2e^{-2\bar{\alpha}z_2},\\
\sigma_{44}^{(2)}&&=\frac{\Gamma_{43}}{\gamma}a_{33}^{(2)}|F|^2e^{-2\bar{\alpha}z_2}=a_{44}^{(2)}|F|^2e^{-2\bar{\alpha}z_2},\\\nonumber
\sigma_{32}^{(2)}&&=\left[\frac{\Omega_c}{\omega+d_{32}}(a_{33}^{(2)}-a_{22}^{(2)})-\frac{a_{21}^{*(1)}}{\omega+d_{32}}\right]|F|^2e^{-2\bar{\alpha}z_2}\\
&&=a_{32}^{(2)}|F|^2e^{-2\bar{\alpha}z_2},
\end{eqnarray}
with $D\equiv|\Omega_c|^2-(\omega+d_{21})(\omega+d_{31})$ and
$D_1\equiv\gamma[(\gamma+\Gamma_{23}+\Gamma_{43})(\gamma+\Gamma_{31})
+\gamma\Gamma_{13}](\omega+d_{32})(\omega+d_{32}^{*})
+2\gamma_{32}[(2\gamma+\Gamma_{43})(\gamma+\Gamma_{31})+\gamma\Gamma_{13}]|\Omega_c|^2$.
$a_{21}^{(1)}$ and $a_{31}^{(1)}$ have been defined in equation
(\ref{1st}).



\begin{thebibliography}{50}
\bibitem{note2}The frequency and wavevector of the probe field are given by
$\omega_p+\omega$ and $k_p+K(\omega)$, respectively. Thus $\omega=0$
corresponds to the center frequency of the probe field
\end{thebibliography}
\end{document}